\documentclass[12pt,a4paper]{article}
\usepackage[dvips]{epsfig}
\usepackage[dvips]{epsfig}
\usepackage{t1enc}
\usepackage[latin1]{inputenc}
\usepackage[english]{babel}
\begin{document}

\thispagestyle{empty}

\title{Evidence for sharper than expected transition between metastable and unstable states}

\author{Dieter W. Heermann and Claudette E. Cordeiro\thanks{Permanent address: Instituto de fisica, 
Universidade Federal Fluminense, 24.210.340-Niteroi-RJ-BRASIL}
         \\ {} \\
           Institut f\"ur Theoretische Physik \\
           Universit\"at Heidelberg \\
           Philosophenweg 19 \\
           D-69120 Heidelberg\\
           and\\ 
           Interdisziplin\"ares Zentrum\\
           f\"ur Wissenschaftliches Rechnen\\
           der Universit\"at Heidelberg \\ {} \\
\vspace {3ex}}

\newpage

\maketitle
\vfill\eject
\begin{abstract}
In mean-field theory, i.e. infinite-range interactions, the transition between metastable and unstable 
states of a thermodynamic system is sharp. The metastable and the unstable states are separated by a 
spinodal curve. For systems with short-range interaction the transition between metastable and unstable 
states has been thought of as gradual. We show evidence, that one can define a sharp border between 
the two regions. We have analysed the lifetimes of states by considering the relaxation trajectories
following a quench. The average lifetimes, as a function of the quench depth into the two-phase region, 
shows a very sharp drop defining a limit of stability for metastable states.
Using the limit of stability we define a line similar to a spinodal in the two-phase region.
\end{abstract}

\vfill\eject

\section{Introduction}

Metastability has traditionally been studied by looking for droplets \cite{Revs1,Revs2}. In this framework density fluctuations are 
produced from the initially homogeneous matrix following a sudden quench from a stable equilibrium state to a state 
underneath the coexistence curve. After a time-lag a quasi-equilibrium is established and a size distribution 
for the fluctuations (droplets) develops, yielding droplets up to a critical size \cite{Becker,Gunton1, Gunton2}.  

Behind this approach to metastability is a geometric interpretation, fostered by experimental
observations of droplets~\cite{Hein}. The experimentally observed droplets are, however, in size much larger than the
critical droplets so that the evidence for nucleation is indirect \cite{Strey}. Also in computer simulations
one can follow the development of the droplet \cite{Stauffer}. Close  to the coexistence curve the droplet picture seems
to hold to a good degree~\cite{Heermann-Klein-Stauffer}, while deeper into the two-phase region considerable doubt has been raised as to
the validity of the nucleation theory \cite{Heermann-Klein}.

To take the point of view of density fluctuations~\cite{Cahn} is by no means the only possible. Here we want to follow a different
point of view. What we want to consider are the fluctuations of a macroscopically available observable and the ensemble
of these fluctuations~\cite{Olivieri1}. To be specific and to link to the simulation results that we present below, consider a
non-conserved order parameter. We start out from a given equilibrium state and suddenly bring the system
into a non-equilibrium state underneath the coexistence curve. The order parameter will follow this change.
Again, after some time--lag the order parameter will fluctuate around some quasi-equilibrium value and suddenly
change to the stable equilibrium state. We now consider the ensemble of such trajectories of the order-parameter.
We avoid to take an average of the order-parameter in the non-equilibrium situation so as not wash out
any of the abrupt changes that take place while the system seeks its way out of the quasi-equilibrium state
into the stable equilibrium state. For each of the trajectories we define a lifetime and consider
the distribution of the lifetimes and also consider the average lifetime! We show below that the average
lifetime dramatically decreases at particular values depending on the imposed parameters giving rise
to a possible dynamic definition of a spinodal. 

So far the existence of a spinodal in the sense of a line of second order phase transition inside
the two-phase region has been denied for systems with short range interactions.  Only for systems with
infinite range interaction (mean field theory) this concept has been thought to make sense~\cite{Revs2}. The concept 
is based on the existence of a free energy and is purely static. Here we side step this and look at 
metastability in a dynamical way and find a clear distinction between metastable and unstable states.

\section{Methodology}

The trajectories discussed above can be obtained naturally using the Ising model 
together with the Monte Carlo Method~\cite{Heermann1,Kalos,Binder-Heermann}. The Monte Carlo method defines the 
transition probabilities between states starting from an initial state.

The  Hamiltonian of the Ising model for a simple cubic lattice $L^3$ is defined by the 

\begin{eqnarray}
{\cal H}_{\mbox{Ising}}(s) = - J \sum_{<ij>}s_is_j\quad + h\sum_{i}s_i, \qquad
s_{i} = \pm 1
\end{eqnarray}

\noindent where $\left<ij\right>$  are nearest-neighbour pairs of lattice sites. The exchange 
coupling $J$ is restricted in our case to be positive (ferromagnetic). 
$s_i$ is called a spin and the sum over all lattice sites is the magnetization $M$ 
(we define $m$ to be the magnetization per spin). $h$ is a dimensionless magnetic field. 

The dynamics of the system is specified by the transition probabilities of
a Markov chain that establishes a path through the available phase space. 
Here we use a  Monte Carlo algorithm~\cite{Heermann1,Kalos,Binder-Heermann} 
to generate Markov Chains.  We used the Metropolis transition probabilities

\begin{equation}
            P(s_i \rightarrow s_i') = \min \{1,\exp (-\Delta {\cal H} )\} \quad .
\end{equation}

\noindent Time $t$ in this context is measured in Monte Carlo steps per spin. One
Monte Carlo step (MCS) per lattice site, i.e.\ one sweep through the
entire lattice, comprises one time unit. 
Neither magnetization nor energy
are preserved in the model which makes possible to compute both quantities as a function
of temperature, applied field and time.

Typically we started the simulation runs with a magnetization of $-1$, i.e. in equilibrium
and a predefined starting point for the random numbers for the Monte Carlo process. 
We then turn on the temperature and the applied field. This brings us instantly below the
coexistence curve into the two-phase region. Due to the transitions induced by the Metropolis
transition probabilities the magnetization (and the energy as well) develops in time and yields a trajectory
of magnetization (energy) values

\begin{equation}
m(0), m(1), ..., m(n)\quad.
\end{equation}

The magnetization was traced to a specific number of Monte Carlo steps. After the prescribed
number of steps were reached, the system was brought back to the magnetization $-1$ and
a new quench performed with a new and different starting point for the random numbers.
The statistics was gathered over these trajectories and quenches with each trajectory kept on file.

\section{Relaxation paths and their statistics}

\begin{figure}[ht]
\begin{center}
\epsfig{file=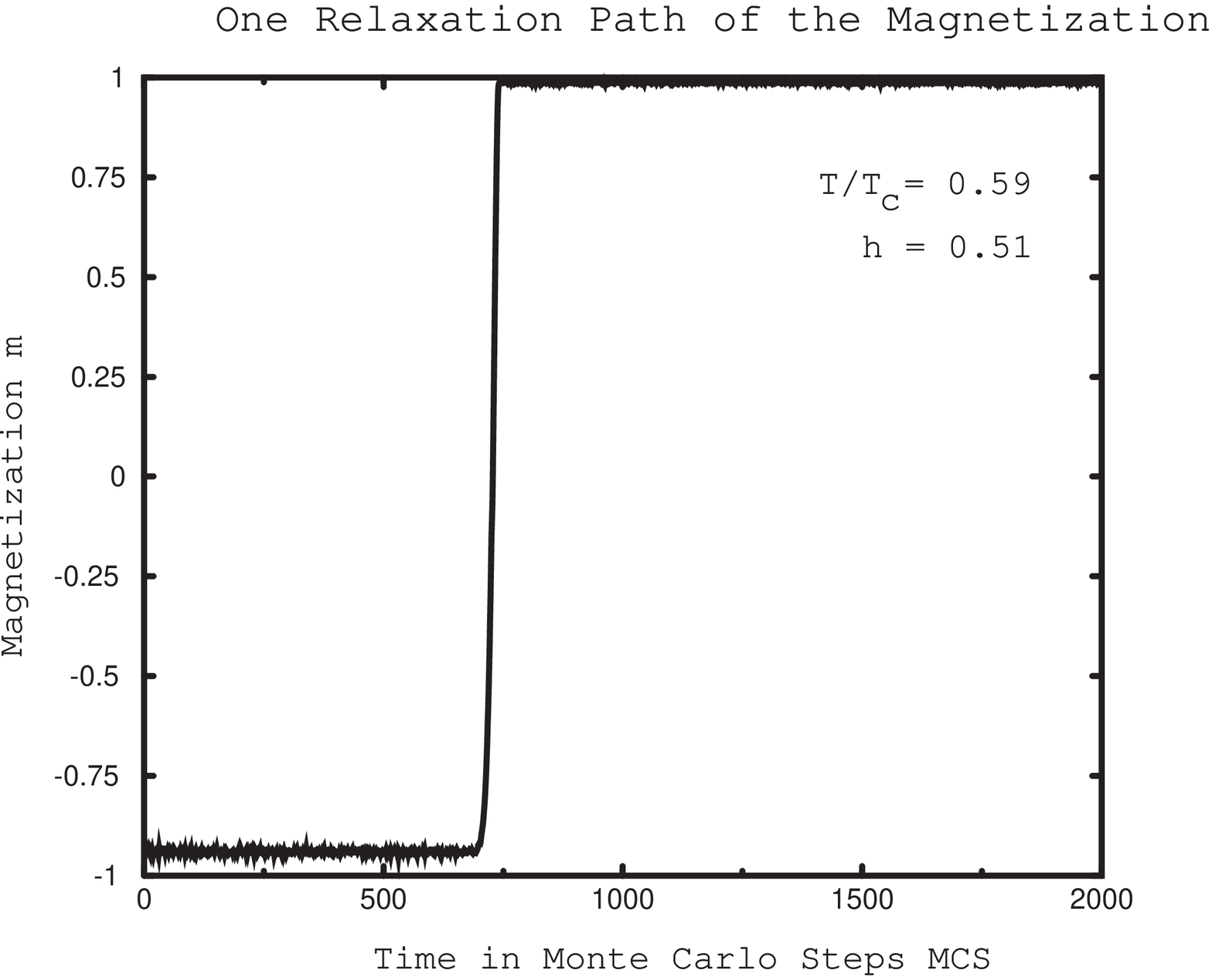,width=1\textwidth}\\
\caption{\label{fig:path} The figure shows one example of relaxation path. Here we display the
magnetization as a function of time following the instant change in the applied magnetic
field $h$ opposite to the magnetization. The system suddenly leaves the quasi-stable state
for the stable equilibrium state corresponding to the temperature and applied field.}
\end{center}
\end{figure}

Let us for the moment fix the temperature ($T=0.59$) and analyse the distribution of lifetimes. 
Figure~\ref{fig:path} shows a typical relaxation path. After fluctuating around some value the magnetization drops sharply. 
At this point we say that the lifetime of the quasi-stable state has been reached. After the sharp drop the 
magnetization reached quickly its value in equilibrium on the other side of the phase diagram.

\begin{figure}[ht]
\begin{center}
\epsfig{file=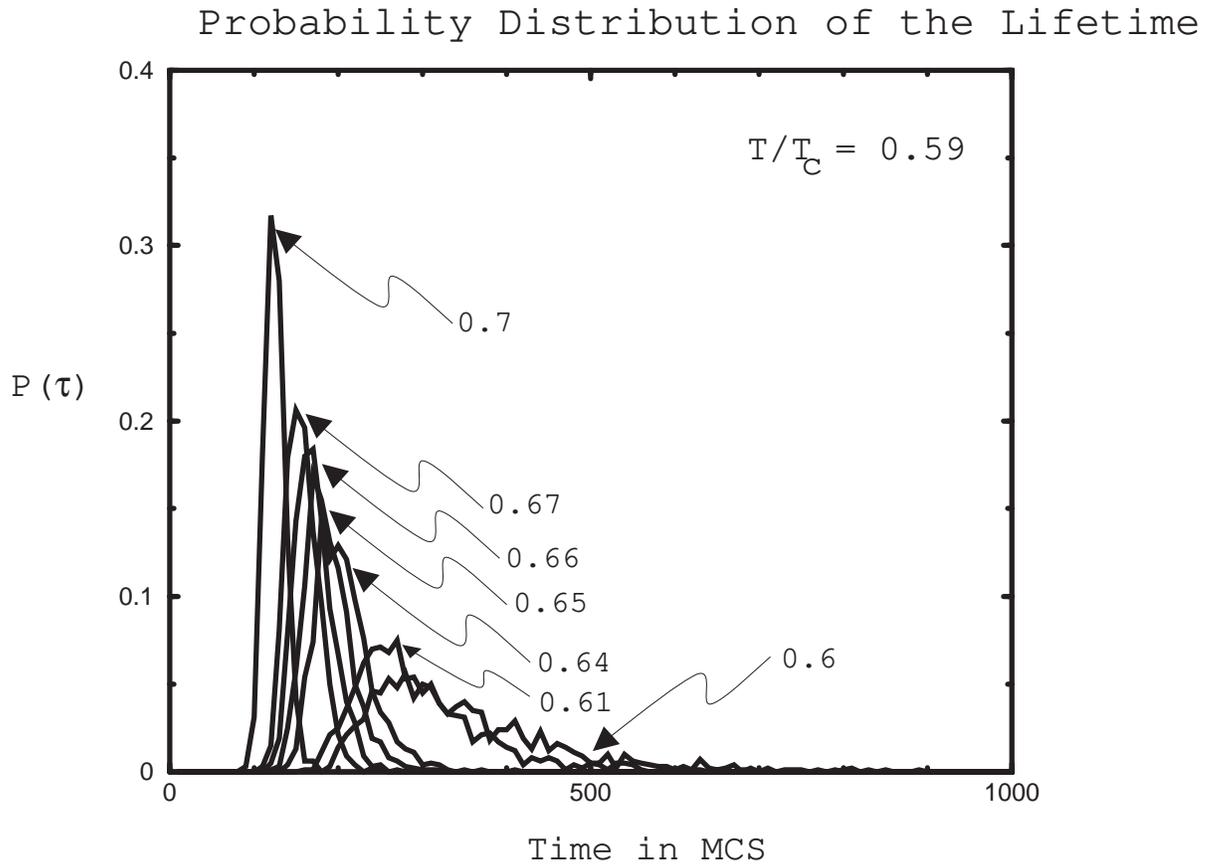,width=1\textwidth}\\
\caption{\label{fig:lifetime-dist} The figure shows the distribution of lifetimes for various applied
fields, i.e. quench-depths. Note that the distribution widens and flattens out as we get closer
to the coexistence curve (h=0). The distribution sharpens as the lifetime gets smaller, i.e. we
quench deeper into the two-phase region!}
\end{center}
\end{figure}

For every quench depth, i.e., applied magnetic field, we can compile a statistics on the lifetimes. This is 
shown in Figure~\ref{fig:lifetime-dist}. Plotted there is the probability for the occurrence of a specific lifetime as a function 
of the quench depth.  Note first that the distribution of lifetimes gets narrower as we quench deeper into the 
two-phase region. Closer to the coexistence curve, i.e. for shallow quenched, the average lifetime increases 
but the width of the distribution broadens. This is of particular interest for experiments. There we also should 
expect a huge variety of lifetimes after a quench making it difficult to interpret the data. Even a quench close 
to the coexistence curve can lead to a very short lifetime and thus to an interpretation that the state may have been 
unstable. In passing we note that the distribution is not Poissonian.

\begin{figure}[ht]
\begin{center}
\epsfig{file=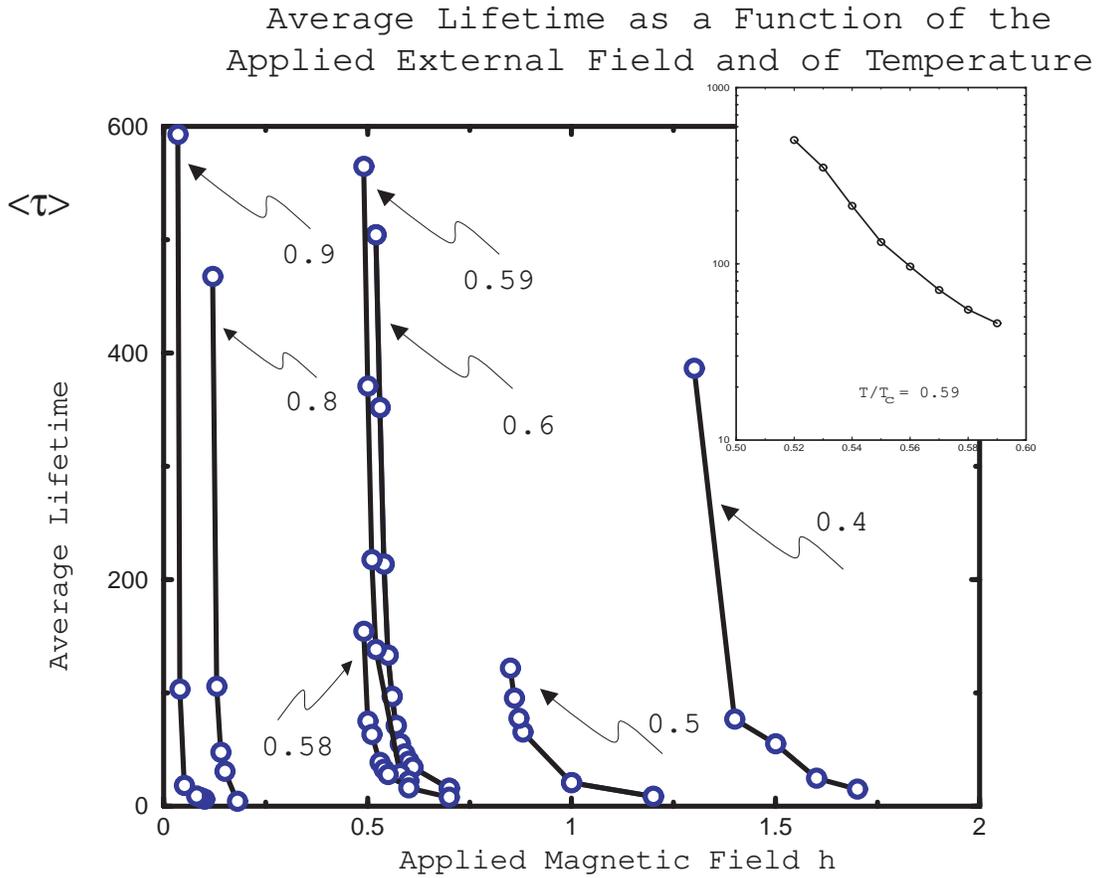,width=1\textwidth}\\
\caption{\label{fig:lifetime} The average lifetime drops sharply as a function of the applied field,
i.e. quench-depth. Shown are the results for various temperatures. The inset shows that the
drop is exponential making it possible to define a reasonably sharp boundary between an
an observable lifetime (metastable state) and those that are inherently unstable.}
\end{center}
\end{figure}

We can plot the average lifetime, at fixed temperature and plot this as a function of the quench depth. 
This is shown in Figure~\ref{fig:lifetime}. While the average lifetime of the states close to the coexistence curve must be very 
large, we expect the lifetime to drop as we quench deeper into the two-phase region. The plot shows that the 
average lifetime decreases rapidly at some  "critical quench depth". The inset in Figure~\ref{fig:lifetime} shows that the average 
lifetime decreases exponentially fast close to that field and then saturates. This saturation depends on the 
precise nature of the Monte Carlo move but must always be present since it always takes a minimum number of 
steps to proceed from one state to another.
 
Clearly, the location of the "limit of stability"-line depends on the kind of Monte Carlo moves that we implement. 
But, the line will always be present for all local move algorithms. For all natural systems we expect this result 
to carry over. Indeed for a pure $CO_2$-system a similar result has been seen~\cite{Wang}. 
There the stability line for states in 
the two-region was also much closer to the coexistence curve then expected.

\section{Finite-Size Effects}

\begin{figure}[ht]
\begin{center}
\epsfig{file=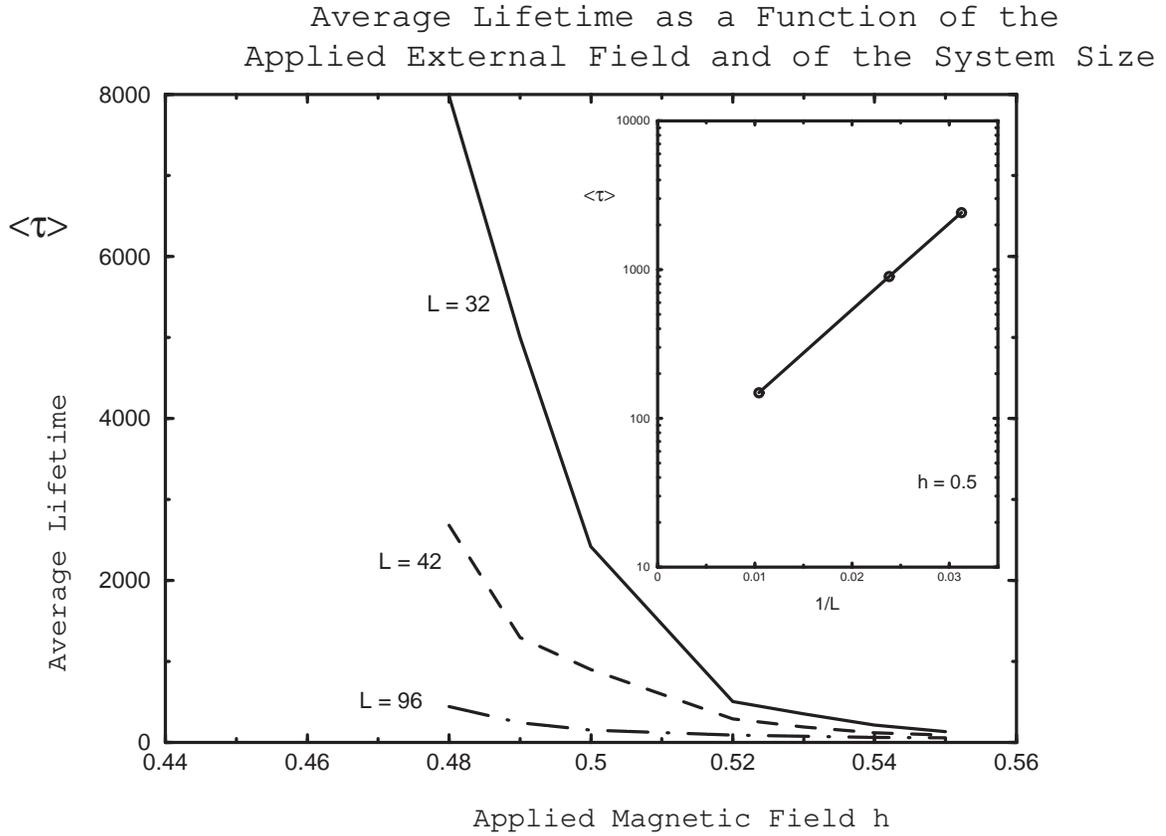,width=1\textwidth}\\
\caption{\label{fig:finite_size} Finite-size effect for the average lifetime. Shown are the results for
the linear system sizes $L=32$, $L=42$  and $L=96$ at the temperature $T/T_c = 0.58$. 
The finite-size effect shows that the pseudo-spinodal will shift closer to the coexistence curve.}
\end{center}
\end{figure}

Figure~\ref{fig:finite_size}  shows the finite-size effects for the average lifetime. 
We have performed simulations with the linear system sizes of $L=32$, $L=42$ and $L=96$ for 
the temperature $T/T_c = 0.58$. First we note that the {\em pseudo spinodal} shifts towards 
the coexistence with increasing system size.  A simple extrapolation gives 
a 30\% shift with respect to the $L=32$ result. We further note the decrease in the average 
lifetime for those quench depths that extend beyond the {\em pseudo spinodal} point. Similar
finite-size effects have also been seen by Novotny et. al.~\cite{Rikvold,Novotny}.

The inset in Figure~\ref{fig:finite_size} shows the extrapolation of the value of the limit 
of stability to the infinite system size.

\begin{figure}[ht]
\begin{center}
\epsfig{file=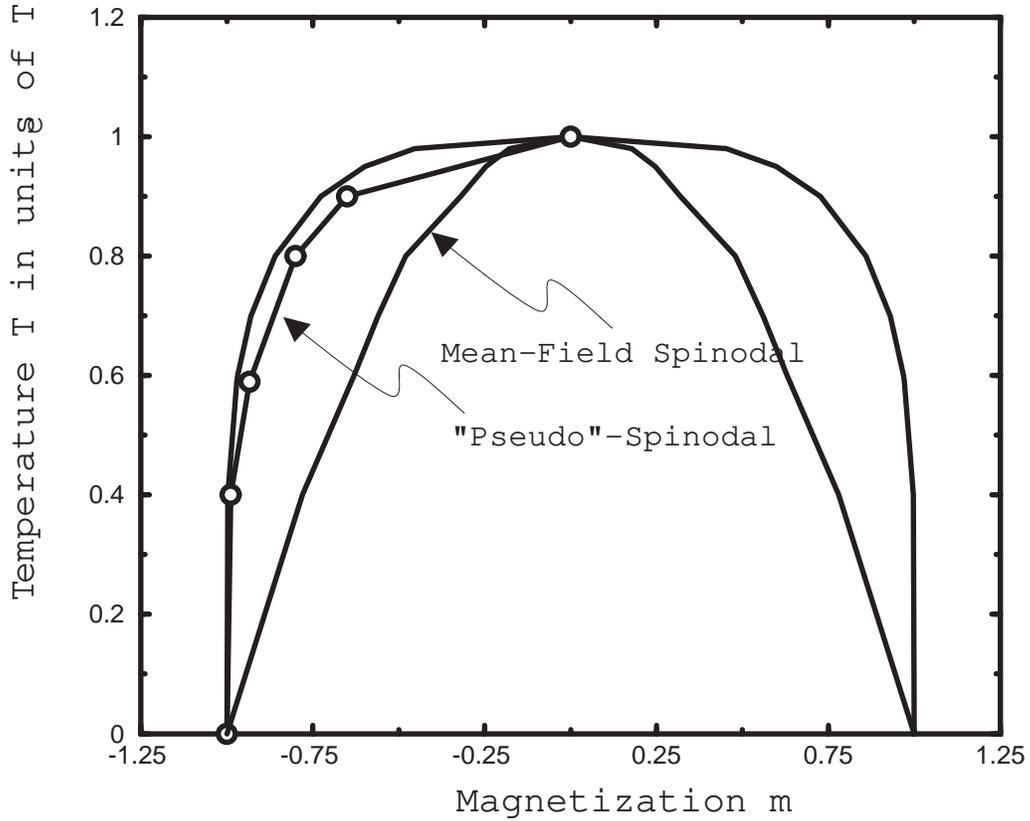,width=1\textwidth}\\
\caption{\label{fig:ab-phase} The figure shows the resulting phase diagram. The phase diagram 
includes the well-known coexistence curve and the mean spinodal curve. The transition line
between metastable and unstable states (pseudo spinodal) is marked by the open circles. Note
that the metastable region is extremely small.}
\end{center}
\end{figure}

In Figure~\ref{fig:ab-phase} we have compiled the results of our findings in a phase
diagram. The new limit of stability is far away from the mean field spinodal curve. 

\section{Conclusions}

Relaxation paths and their statistics yield a fresh outlook on metastability. 
Here the emphasis is on the dynamics of the single events and not on the average behaviour 
of the system. Our point of view is that averaging in the non-equilibrium situation can 
wash away or hide the phenomenon. Here we have shown that the statistics of the relaxation 
paths can lead to a sharp distinction between meta- and unstable states. 

A surprising finding is that the distribution of the lifetime broadens significantly. 
Opposite to the intuition we found that close to coexistence curve the distribution is 
broad and very flat. Close to the limit of stability the distribution is very 
narrow and sharply peaked.

Clearly further work needs to be done, cementing the evidence for the sharp border, in the sense
defined above. As pointed out above, simulations using different transition probabilities 
would be helpful to investigate the dependence of this dynamic phenomenon on the imposed 
dynamics, at least for those systems that do not have an intrinsic dynamics.

It would also be of interest to perform molecular dynamics simulations of systems using
realistic interaction potential. There the question would also be how far the metastable 
region really extends and compare this to experimental evidence for nucleation or spinodal 
decomposition in the light of the relaxation trajectories.

It is also unclear at the moment how to extend the definition to systems with a 
conserved order parameter.

\section{Acknowledgment}

This work was funded by a grant from BASF AG and a grant from the BMBF. 
C.E.C would like to thank the DAAD and FAPERJ for financial support.
We are also
very grateful for discussions with K. Binder, E. H\"adicke, B. Rathke, R. Strey and D. Stauffer.


\newpage

\end{document}